# Modulation theory in $\mathcal{PT}$-symmetric magnetic metamaterial arrays in the continuum limit

Danhua Wang and Alejandro B. Aceves

*Department of Mathematics, Southern Methodist University, Dallas, Texas 75275, USA*

We present results on the dynamics of split-ring dimers having both gain and loss in one-dimensional nonlinear parity-time- ($\mathcal{PT}$-)symmetric magnetic metamaterials. For the long-wave (continuum) limit approximation and in the weakly nonlinear limit, we show analytic results on the existence of gap soliton solutions and on symmetry-breaking phenomenon at a critical value of the gain or loss term.

## I. INTRODUCTION

The study of $\mathcal{PT}$-symmetric systems has received considerable attention as it may provide a new framework for a number of applications. In the simplest possible optical coupled systems, the experimental observation of $\mathcal{PT}$ symmetry [1] builds on managing the balance of gain and loss in an otherwise conservative nonlinear system. Such symmetry breaking can have new all optical applications such as a unidirectional optical valve [2]. Perhaps because controlling gain or loss is less difficult in discrete photonic systems than in bulk, much work on $\mathcal{PT}$-symmetric systems considers optical lattices [3], or for example binary arrays of optical waveguides [4]. From the theoretical point of view, for finite lattice systems (dimers, trimer, oligomers) a classical dynamical systems approach can be used to search for solutions, their stability, and bifurcation properties [5,6]. For extended (ideally infinite) systems, the building blocks are discrete solitons of a modified discrete nonlinear Schrödinger equation (d-NLSE), found for example as a bifurcation from the anticontinuum limit [5] whose persistence and eventual symmetry breaking behavior in terms of the gain-loss parameter reveals the $\mathcal{PT}$-symmetry breaking behavior. In most instances, assuming the scaling validates the assumption, a continuum approximation is used leading to the nonlinear Schrödinger equation with a linear complex potential whose real part is even and imaginary part is odd. In such case questions of existence [7,8] and stability [9] can be performed as more analytical tools are available. When the basic element of an array is a coupled waveguide element with one guide having linear gain and the other one linear loss of equal strength, and if one applies a continuum approximation one arrives at a coupled nonlinear Schrödinger equation (c-NLSE) [10] where existence and stability analysis is the same as classical work on similar c-NLSE systems.

With the development of novel engineered metamaterials, it is now conceivable to tailor dielectric properties to achieve $\mathcal{PT}$ symmetry and control its symmetry breaking in ways described by models based on the d-NLSE, it is also the case that these ideas have been exploited in electronic circuits where experimental conditions are more accessible. Take, for example, recent work on LRC circuits with $\mathcal{PT}$ symmetry [11]. Another possibility is by use of split-ring resonators (SRRs) and SRRs arrays [12,13].

As stated before, in the simplest form, proof of principle of symmetry-breaking dynamics in any of these optically or electronically based models is present as a universal bifurcation past a threshold of the symmetric nonconservative (gain-loss) perturbation of an otherwise conservative system. Here, we pay attention to recent work on a $\mathcal{PT}$-dimer chain [12] shown schematically in Fig. 1. In this work, the authors demonstrate numerically that discrete breathers in such a dimer chain are generic though their long term stability is compromised when the balance between gain and loss is not exact. It is not clear though if a symmetry breaking bifurcation exists. Noticing that most relevant parameters in the array are small, in our work we benefit from it to develop a weakly nonlinear theory and derive amplitude equations in both the discrete and the continuum approximation. The results we present here correspond to the case where the continuum limit applies. In doing so, the sections that follow show, in particular, that we can analytically predict the existence of gap solitonlike solutions in the absence of gain or loss. Similarly, we are able to predict the bifurcation value of the loss-gain parameter about which localized modes cease to exist. The conclusions summarize our results and suggest possible extensions.

## II. DISCRETE MODEL

In the equivalent circuit model picture, extended for the $\mathcal{PT}$ dimer chain, the dynamics of the charge $q_n$ in the capacitor of the $n$th SRR is governed by [12]

$$\lambda'_M \ddot{q}_{2n} + \ddot{q}_{2n+1} + \lambda_M \ddot{q}_{2n+2} + \lambda'_E q_{2n} + q_{2n+1} + \lambda_E q_{2n+2}$$
$$= \varepsilon_0 \sin(\Omega t) - \alpha q_{2n+1}^2 - \beta q_{2n+1}^3 - \gamma \dot{q}_{2n+1},$$
$$\lambda_M \ddot{q}_{2n-1} + \ddot{q}_{2n} + \lambda'_M \ddot{q}_{2n+1} + \lambda_E q_{2n-1} + q_{2n} + \lambda'_E q_{2n+1}$$
$$= \varepsilon_0 \sin(\Omega t) - \alpha q_{2n}^2 - \beta q_{2n}^3 + \gamma \dot{q}_{2n}, \quad (1)$$

where $\lambda_M$, $\lambda'_M$ and $\lambda_E$, $\lambda'_E$ are the magnetic and electric interaction coefficients, respectively, between nearest neighbors, $\alpha$ and $\beta$ are nonlinear coefficients, $\gamma$ is the gain or loss coefficient ($\gamma > 0$), $\varepsilon_0$ is the amplitude of the external driving voltage, while $\Omega$ and $t$ are the driving frequency and temporal variable, respectively, normalized to $\omega_0 = 1/\sqrt{LC_0}$ and $\omega_0^{-1}$, respectively, with $C_0$ being the linear capacitance. In the following, we only consider that the relative orientation of the SRRs in the chain is such that the magnetic coupling dominates, while the electric coupling can be neglected, i.e., $\lambda_E = \lambda'_E = 0$.

Substituting $A_m = q_{2n+1}$ and $B_m = q_{2n}$ into Eqs. (1), we obtain

$$\ddot{A}_m + A_m = -\lambda'_M \ddot{B}_m - \lambda_M \ddot{B}_{m+1} + \varepsilon_0 \sin(\Omega t) - \alpha A_m^2 - \beta A_m^3 - \gamma \dot{A}_m,$$
$$\ddot{B}_m + B_m = -\lambda_M \ddot{A}_{m-1} - \lambda'_M \ddot{A}_m + \varepsilon_0 \sin(\Omega t) - \alpha B_m^2 - \beta B_m^3 + \gamma \dot{B}_m. \quad (2)$$

By assuming the parameter values in [12], all the terms in the right-hand side are small since $|\lambda'_M|, |\lambda_M|, |\varepsilon_0|, |\alpha|, |\beta|, |\gamma| \ll 1$. If we first start by studying the dispersion properties in the linear regime, with no forcing for which we assume plane-wave solutions, $(A_m(t), B_m(t)) \propto e^{i(mk_x + k_t t)}$, we obtain the linear dispersion relation

$$k_t^2 = \frac{2 - \gamma^2 \pm \sqrt{\gamma^4 - 4\gamma^2 + 4(\lambda_M - \lambda'_M)^2 + 16\cos^2\frac{k_x}{2}\lambda_M\lambda'_M}}{2[1 - (\lambda_M - \lambda'_M)^2 - 4\cos^2\frac{k_x}{2}\lambda_M\lambda'_M]},$$

which is consistent with $\Omega_\kappa^2$ in [12] where $k_x = 2\kappa$ is the wave vector. Observe that to first approximation $k_t \approx \pm 1$ corresponding to ignoring the right-hand side of Eq. (2). This serves as the basis to develop the weakly nonlinear modulation theory which assumes $A_m = u_m(\tau)e^{it+k_m x} + u_m^*(\tau)e^{-(it+k_m x)} + \epsilon u_m^{(2)}$, $B_m = v_m(\tau)e^{it+k_m x} + v_m^*(\tau)e^{-(it+k_m x)} + \epsilon v_m^{(2)}$, where $\tau = \epsilon t$, $0 < \epsilon \ll 1$ is a slow time scale. Then the slowly varying amplitude equations, in the absence of forcing, satisfy the equations

$$2i\frac{du_m}{d\tau} = \bar{\lambda}_M v_m + \bar{\lambda}_M v_{m+1}e^{ik_x} - 3\bar{\beta}|u_m|^2 u_m - i\bar{\gamma}u_m, \quad 2i\frac{dv_m}{d\tau} = \bar{\lambda}_M u_m + \bar{\lambda}_M u_{m-1}e^{-ik_x} - 3\bar{\beta}|v_m|^2 v_m + i\bar{\gamma}v_m \quad (3)$$

and

$$u_M^{(2)} = -2\bar{\alpha}|u_m|^2 + \frac{\bar{\alpha}}{3}u_m^2 e^{2i(t+k_x m)} + \frac{\bar{\alpha}}{3}u_m^{*2} e^{-2i(t+k_x m)}, \quad v_M^{(2)} = -2\bar{\alpha}|v_m|^2 + \frac{\bar{\alpha}}{3}v_m^2 e^{2i(t+k_x m)} + \frac{\bar{\alpha}}{3}v_m^{*2} e^{-2i(t+k_x m)}.$$

Here all parameters are rescaled as $\mu \to \epsilon \bar{\mu}$.

Rather than analyzing the modulations equations just derived, the rest of the paper deals with the long-wave continuum approximation limit.

### III. CONTINUUM APPROXIMATION

The long-wave limit ($k_x \approx 0$) can be better analyzed if we use the continuum approximation given by the expansions

$$A_{m\pm 1}(z) = u(x,t) \pm u_x(x,t) + \tfrac{1}{2}u_{xx}(x,t) + \cdots, \quad B_{m\pm 1}(t) = w(x,t) \pm w_x(x,t) + \tfrac{1}{2}w_{xx}(x,t) + \cdots,$$

obtaining (as a first-order approximation)

$$u_{tt} + u = \varepsilon_0 \sin(\Omega t) - a_M w_{tt} - \lambda_M w_{ttx} - \alpha u^2 - \beta u^3 - \gamma u_t,$$
$$w_{tt} + w = \varepsilon_0 \sin(\Omega t) - a_M u_{tt} + \lambda_M u_{ttx} - \alpha w^2 - \beta w^3 + \gamma w_t, \quad (4)$$

where $a_M = \lambda'_M + \lambda_M$. Observe that in the linear case if we set $\varepsilon_0 = 0$, $\alpha = 0$, and $\beta = 0$ the dispersion relation of Eqs. (4) reads

$$k_t^2 = \frac{2 - \gamma^2 \pm \sqrt{\gamma^4 - 4\gamma^2 + 4a_M^2 + 4k_x^2\lambda_M^2}}{2[1 - a_M^2 - k_x^2\lambda_M^2]} = \frac{2 - \gamma^2 \pm \sqrt{\gamma^4 - 4\gamma^2 + 4(\lambda_M - \lambda'_M)^2 + 4k_x^2\lambda_M^2 + 16\lambda_M\lambda'_M}}{2[1 - (\lambda_M - \lambda'_M)^2 - k_x^2\lambda_M^2 - 4\lambda_M\lambda'_M]},$$

which corresponds to the first approximation to the longwave limit $k_x \approx 0$ of the discrete model, thus validating the continuum model. Next we want to find regions in terms of $k_x$ and $\gamma$ where $k_t$ is real. We should point out that while we present a detailed picture of the linear dispersion relation, only those regions where $k_x$ is close to zero [case (iii) and Fig. 3] are consistent with the approximation.

(i) If $k_x^2 > \frac{1}{\lambda_M^2}[1 - a_M^2]$, then

$$k_t^2 = \frac{2 - \gamma^2 - \sqrt{\gamma^4 - 4\gamma^2 + 4a_M^2 + 4k_x^2\lambda_M^2}}{2[1 - a_M^2 - k_x^2\lambda_M^2]} > 0 \quad (5)$$

are real for $\gamma > 0$ (see Fig. 2 top).

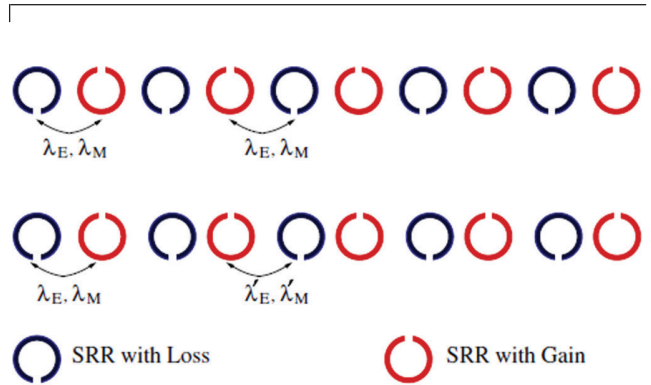

FIG. 1. (Color online) Schematic of a $\mathcal{PT}$ metamaterial (as in Fig. 1 in [12]). Upper panel: all the SRRs are equidistant. Lower panel: the separation between SRRs is modulated according to a binary pattern ($\mathcal{PT}$ dimer chain).

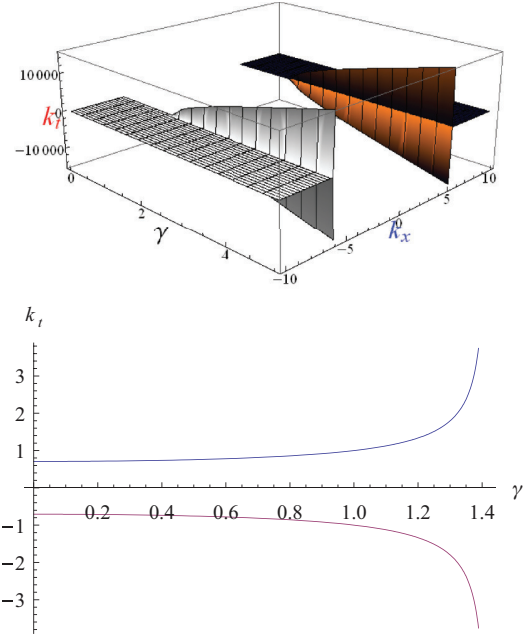

FIG. 2. (Color online) If $\lambda_M = -0.17$ and $\lambda'_M = -0.10$. (a) Linear dispersion relation for regime (i) (5); (b) linear dispersion relation for fixed $k_x = \frac{1}{\|\lambda_M\|}\sqrt{1-a_M^2}$ (6).

(ii) If $k_x^2 = \frac{1}{\lambda_M^2}[1-a_M^2]$, then

$$k_t^2 = \frac{1}{2-\gamma^2} > 0 \qquad (6)$$

are real for $0 < \gamma < \sqrt{2}$ (see Fig. 2 bottom).

(iii) If $\frac{1}{\lambda_M^2}[-\frac{\gamma^4}{4}+\gamma^2-a_M^2] \leqslant k_x^2 < \frac{1}{\lambda_M^2}[1-a_M^2]$, then

$$k_t^2 = \frac{2-\gamma^2 \pm \sqrt{\gamma^4 - 4\gamma^2 + 4a_M^2 + 4k_x^2\lambda_M^2}}{2[1-a_M^2-k_x^2\lambda_M^2]} > 0 \qquad (7)$$

are real for $\sqrt{2-2\sqrt{1-a_M^2}} < \gamma < \sqrt{2}$ (see Fig. 3).

As we can see from Fig. 3, the the presence of the loss-gain parameter opens the gap in around $k_x = 0$ up to where $\gamma = \sqrt{2}$ where there is a full gap.

## IV. WEAKLY NONLINEAR THEORY

Recognizing that most parameters are small, we apply regular perturbation theory on Eqs. (1) and (4) to derive a weakly nonlinear model.

For the sake of simplicity, we rescale all small parameters $\lambda'_M$, $\lambda_M$, $\varepsilon_0$, $\alpha$, $\beta$, $\gamma$ in Eqs. (1) and (4) as $\mu \to \epsilon\mu$, where $\mu$ represents any of these parameters and where $\epsilon$ is small. We describe in some detail the weakly nonlinear theory for the continuum model and for the discrete model we simply present the equivalent outcome.

Let

$$u = u_0 + \epsilon u_1 + \epsilon^2 u_2 + \cdots, \quad w = w_0 + \epsilon w_1 + \epsilon^2 w_2 + \cdots,$$

We have

$$u_{0tt} + u_0 = 0, \quad w_{0tt} + w_0 = 0.$$

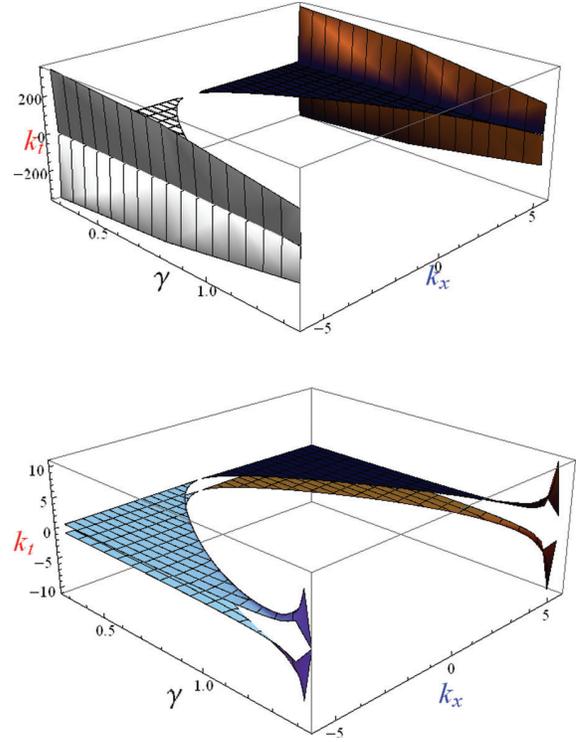

FIG. 3. (Color online) If $\lambda_M = -0.17$ and $\lambda'_M = -0.10$. Linear dispersion relation (7) (sign $\pm$).

$u_0(x,t)$, $w_0(x,t)$ would be the known solution to the uncoupled harmonic-oscillator equation and $u_1(x,t), u_2(x,t), \ldots, w_1(x,t), w_2(x,t), \ldots$ represent the higher-order terms which are found iteratively. One obtains the solutions to leading order as

$$u_0(x,t) = A(x,\tau)e^{it} + A^*(x,\tau)e^{-it},$$
$$w_0(x,t) = B(x,\tau)e^{it} + B^*(x,\tau)e^{-it},$$

where $*$ is the operator of complex conjugate and $\tau = \frac{1}{2}\epsilon|\lambda_M|t$. Substituting in Eqs. (4) and applying the solvability conditions at order $O(\epsilon)$ to remove secular terms in

$$u_{1tt} + u_1 = \left[-i\frac{\partial A}{\partial \tau}e^{it} + i\frac{\partial A^*}{\partial \tau}e^{-it} + \varepsilon_0\sin(\Omega t) - a_M w_{0tt}\right.$$
$$\left. - \lambda_M w_{0ttx} - \alpha u_0^2 - \beta u_0^3 - \gamma u_{0t}\right],$$

$$w_{1tt} + w_1 = \left[-i\frac{\partial B}{\partial \tau}e^{it} + i\frac{\partial B^*}{\partial \tau}e^{-it} + \varepsilon_0\sin(\Omega t) - a_M u_{0tt}\right.$$
$$\left. + \lambda_M u_{0ttx} - \alpha w_0^2 - \beta w_0^3 + \gamma w_{0t}\right] \qquad (8)$$

gives the slowly varying equations for $A, B$.

(i) No resonant forcing ($\Omega \neq 1$),

$$-i\frac{\partial B}{\partial \tau} + \frac{\partial A}{\partial x} = \frac{1}{\lambda_M}[a_M A - 3\beta|B|^2 B + i\gamma B],$$
$$i\frac{\partial A}{\partial \tau} + \frac{\partial B}{\partial x} = \frac{1}{\lambda_M}[-a_M B + 3\beta|A|^2 A + i\gamma A]. \qquad (9)$$

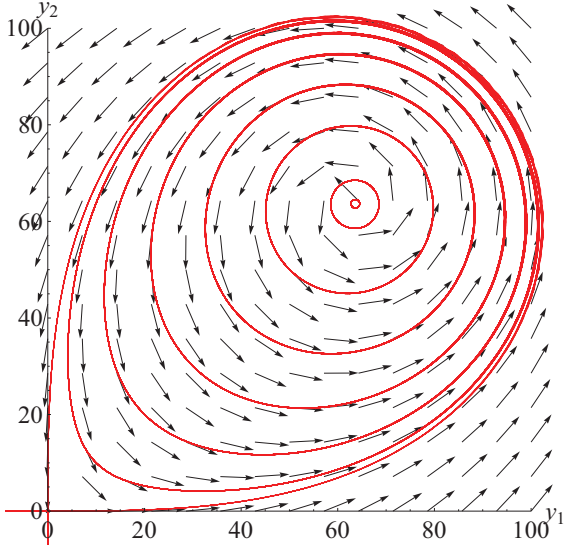

FIG. 4. (Color online) Phase plane of the real ODE system (17) for $a_M = -0.27$, $\lambda_M = -0.17$, and $\beta = 0.002$.

(ii) At resonance $\Omega = 1$,

$$-i\frac{\partial B}{\partial \tau} + \frac{\partial A}{\partial x} = \frac{1}{\lambda_M}\left[a_M A - 3\beta|B|^2 B + i\gamma B - \frac{i}{2}\varepsilon_0\right],$$
$$i\frac{\partial A}{\partial \tau} + \frac{\partial B}{\partial x} = \frac{1}{\lambda_M}\left[-a_M B + 3\beta|A|^2 A + i\gamma A + \frac{i}{2}\varepsilon_0\right]. \quad (10)$$

(iii) Near resonance $\Omega = 1 + \epsilon\omega$,

$$-i\frac{\partial B}{\partial \tau} + \frac{\partial A}{\partial x} = \frac{1}{\lambda_M}\left[a_M A - 3\beta|B|^2 B + i\gamma B - \frac{i}{2}\varepsilon_0 e^{i\omega\tau}\right],$$
$$i\frac{\partial A}{\partial \tau} + \frac{\partial B}{\partial x} = \frac{1}{\lambda_M}\left[-a_M B + 3\beta|A|^2 A + i\gamma A + \frac{i}{2}\varepsilon_0 e^{i\omega\tau}\right]. \quad (11)$$

and the $O(\epsilon)$ corrections for the nonresonant $[\frac{\varepsilon_0}{1-\Omega^2} = O(1)]$ case being,

$$u_1(x,t) = \frac{\varepsilon_0}{2(1-\Omega^2)i}(e^{i\Omega t} - e^{-i\Omega t}) - 2\alpha|A(x)|^2$$
$$+ \frac{\alpha}{3}[A(x)^2 e^{2it} + A^*(x)^2 e^{-2it}]$$
$$+ \frac{\beta}{8}[A(x)^3 e^{3it} + A^*(x)^3 e^{-3it}],$$
$$w_1(x,t) = \frac{\varepsilon_0}{2(1-\Omega^2)i}(e^{i\Omega t} - e^{-i\Omega t}) - 2\alpha|B(x)|^2$$
$$+ \frac{\alpha}{3}[B(x)^2 e^{2it} + B^*(x)^2 e^{-2it}]$$
$$+ \frac{\beta}{8}[B(x)^3 e^{3it} + B^*(x)^3 e^{-3it}]. \quad (12)$$

### A. Gap soliton solutions

Equations (9), (10), and (11) above belong to the family of systems of strongly coupled modes with a gap in the dispersion relation, with numerous examples emerging the past 25 years. In [14,15], in their study of Bragg grating solitons, it was first noticed that solutions of the integrable Thirring model could be extended to similar nonintegrable systems. This approach has been recently applied in a model of a delay line in a dual core photonic crystal fiber [16] and in binary arrays [17]. Similarly here, fully time-dependent gap solitonlike solutions can be obtained in the nonresonant case [Eqs. (9)] for $\gamma = 0$. Following the set of transformations in [17], and for $\gamma = 0$, one can find exact solitary wave solutions. Namely, if we write solutions of Eqs. (9) in the form

$$B(x,\tau) = \frac{1}{2i}[K_1 g_1(\xi) - i K_2 g_2(\xi)]\exp(i\psi \cos Q), \quad (13)$$

$$A(x,\tau) = \frac{1}{2}[K_1 g_1(\xi) + i K_2 g_2(\xi)]\exp(i\psi \cos Q), \quad (14)$$

$$\xi = \frac{x + v\tau}{\sqrt{1-v^2}}, \quad \psi = \frac{vx + \tau}{\sqrt{1-v^2}},$$

$$K_1 = \left(\frac{1+v}{1-v}\right)^{\frac{1}{4}}, \quad K_2 = \left(\frac{1-v}{1+v}\right)^{\frac{1}{4}},$$

with $g_{1,2}$ two arbitrary complex functions, $-1 \leqslant v \leqslant 1$ and $0 \leqslant Q \leqslant \pi$. Substituting (13) and (14) into Eqs. (9) gives

$$-\dot{g}_1 + ig_1 \cos Q - \frac{ia_M}{\lambda_M}g_2 + \frac{3\beta}{4i\lambda_M}$$
$$\times (K_1^4|g_1|^2 g_1 + 2|g_2|^2 g_1 - g_2^2 g_1^*) = 0,$$
$$\dot{g}_2 + ig_2 \cos Q - \frac{ia_M}{\lambda_M}g_1 + \frac{3\beta}{4i\lambda_M}$$
$$\times (K_2^4|g_2|^2 g_2 + 2|g_1|^2 g_2 - g_1^2 g_2^*) = 0.$$

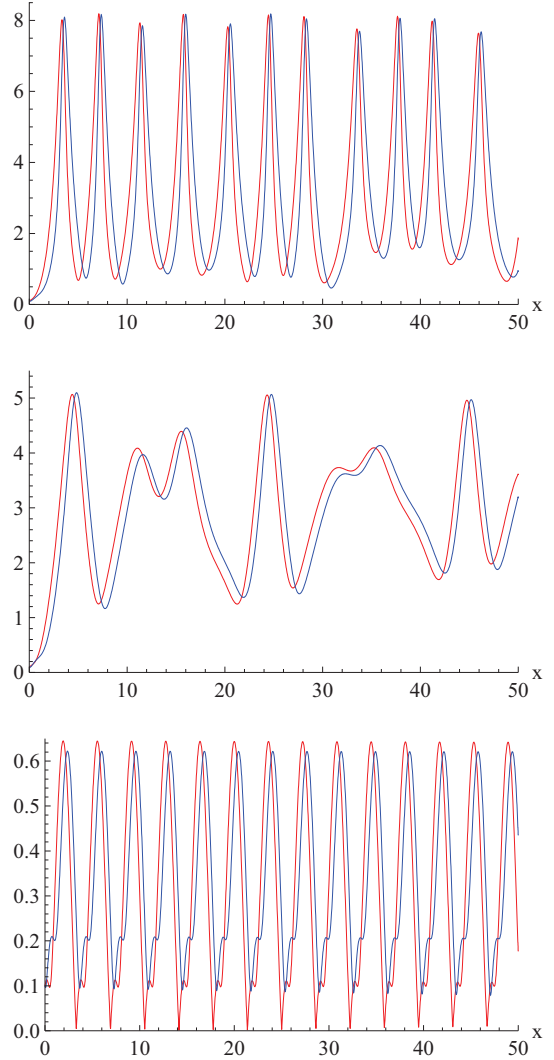

FIG. 5. (Color online) Numerical solutions $|A(x)|$ (red) and $|B(x)|$ (blue) of (10) for $\varepsilon_0 = 0.1$, $a_M = -0.27$, $\lambda_M = -0.17$, and $\beta = 0.002$. Top: $\gamma = 0.2$; middle: $\gamma = 0.27$; bottom: $\gamma = 0.4$.

These equations imply the invariant $P = |g_1|^2 - |g_2|^2$. In the case $P = 0$, we have $|g_1|^2 = |g_2|^2$ and $g_{1,2}(\xi) = f(\xi) \exp[i\theta_{1,2}(\xi)]$. Therefore, by $\mu = f^2$ and $v = \theta_1 - \theta_2$, we obtain

$$\dot{\mu} = -\frac{\partial H}{\partial v}, \quad \dot{v} = \frac{\partial H}{\partial \mu},$$

$$H = 2\mu \left(-\frac{a_M}{\lambda_M} \cos v + \cos Q\right)$$
$$- \frac{3\beta}{4\lambda_M} \mu^2 \left(\frac{K_1^4}{2} + \frac{K_2^4}{2} + 2 - \cos(2v)\right). \quad (15)$$

Equations (15) represent a one-dimensional integrable Hamiltonian system from which solitary wave solutions are obtained [17]. It would be of interest if by use of Hamiltonian perturbative methods [18], such solutions are stable and persist for small but nonzero $\gamma$ and small nonzero $\gamma, \epsilon_0$ for the resonant case. We leave this for future work.

### B. Stationary solutions

In this section, we consider the existence of stationary solutions of the amplitude equations and in particular we show a bifurcation for the nonresonant case, which is the first case we consider. To do so, letting $y_1 = |A|^2, y_2 = |B|^2, y_3 = (AB^* + A^*B), y_4 = i(AB^* - A^*B)$, we could rewrite (9) as a real ODE system

$$\frac{dy_1}{dx} = \frac{1}{\lambda_M}(2a_M y_1 - 3\beta y_2 y_3 - \gamma y_4),$$
$$\frac{dy_2}{dx} = \frac{1}{\lambda_M}(-2a_M y_2 + 3\beta y_1 y_3 + \gamma y_4),$$

$$\frac{dy_3}{dx} = \frac{6\beta}{\lambda_M}(y_1^2 - y_2^2),$$
$$\frac{dy_4}{dx} = \frac{2\gamma}{\lambda_M}(y_1 - y_2). \quad (16)$$

If $\gamma = 0$ in Eqs. (16), there is an invariant

$$y_3 = \frac{3\beta}{2a_M}(y_1^2 + y_2^2).$$

Then the real ODE system (16) becomes a 2D integrable system:

$$\frac{dy_1}{dx} = \frac{1}{\lambda_M}\left[2a_M y_1 - \frac{9\beta^2}{2a_M} y_2(y_1^2 + y_2^2)\right],$$
$$\frac{dy_2}{dx} = \frac{1}{\lambda_M}\left[-2a_M y_2 + \frac{9\beta^2}{2a_M} y_1(y_1^2 + y_2^2)\right]. \quad (17)$$

The Jacobian matrix of (17) is

$$\begin{pmatrix} \frac{1}{\lambda_M}(2a_M - \frac{9\beta^2}{a_M} y_1 y_2) & -\frac{9\beta^2}{2\lambda_M a_M}(y_1^2 + 3y_2^2) \\ \frac{9\beta^2}{2\lambda_M a_M}(3y_1^2 + y_2^2) & -\frac{1}{\lambda_M}(2a_M - \frac{9\beta^2}{a_M} y_1 y_2) \end{pmatrix}.$$

For Eqs. (17), there are two equilibrium points: one is a saddle $(0,0)$ at which the eigenvalues are $\lambda_{1,2} = \pm 2|\frac{a_M}{\lambda_M}|$ and the corresponding eigenvectors are $(1,0)^T, (0,1)^T$. And another one is a center at $(\frac{\sqrt{2}}{3}|\frac{a_M}{\beta}|, \frac{\sqrt{2}}{3}|\frac{a_M}{\beta}|)$ with eigenvalues $\lambda_{1,2} = \pm 4|\frac{a_M}{\lambda_M}|i$. The phase plane of this integrable case is shown as in Fig. 4.

Interestingly, as it is the case of many $\mathcal{PT}$ systems, this symmetry reflects in an invariance. Here, if $\gamma > 0$ in Eqs. (16),

$$y_3 = \frac{3\beta}{2a_M}\left(y_1^2 + y_2^2 + \frac{1}{2} y_4^2\right)$$

in a constant. Then the real ODE system (16) can be reduced into the three-dimensional system,

$$\frac{dy_1}{dx} = \frac{1}{\lambda_M}\left[2a_M y_1 - \frac{9\beta^2}{2a_M} y_2\left(y_1^2 + y_2^2 + \frac{1}{2} y_4^2\right) - \gamma y_4\right],$$
$$\frac{dy_2}{dx} = \frac{1}{\lambda_M}\left[-2a_M y_2 + \frac{9\beta^2}{2a_M} y_1\left(y_1^2 + y_2^2 + \frac{1}{2} y_4^2\right) + \gamma y_4\right],$$
$$\frac{dy_4}{dx} = \frac{2\gamma}{\lambda_M}(y_1 - y_2), \quad (18)$$

for which the Jacobian matrix of the $\underline{0}$ solution is

$$\begin{pmatrix} \frac{1}{\lambda_M}(2a_M - \frac{9\beta^2}{a_M} y_1 y_2) & -\frac{9\beta^2}{2\lambda_M a_M}(y_1^2 + 3y_2^2 + \frac{1}{2} y_4^2) & -\frac{1}{\lambda_M}(\frac{9\beta^2}{2a_M} y_2 y_4 + \gamma) \\ \frac{9\beta^2}{2\lambda_M a_M}(3y_1^2 + y_2^2 + \frac{1}{2} y_4^2) & -\frac{1}{\lambda_M}(2a_M - \frac{9\beta^2}{a_M} y_1 y_2) & \frac{1}{\lambda_M}(\frac{9\beta^2}{2a_M} y_1 y_4 + \gamma) \\ \frac{2\gamma}{\lambda_M} & -\frac{2\gamma}{\lambda_M} & 0 \end{pmatrix}.$$

The eigenvalues are $\lambda_1 = 0$ and $\lambda_{2,3} = \pm\frac{2}{|\lambda_M|}\sqrt{a_M^2 - \gamma^2}$. Then the equilibrium point is a saddle when $0 < \gamma < |a_M|$, while it is a center when $\gamma > |a_M|$. So we have proven there is a bifurcation from saddle to center at $\gamma_c = |a_M|$.

We confirm the previous result from the full system where the Jacobian matrix of the real ODE system (16) for the equilibrium solution at the origin is

$$\begin{pmatrix} \frac{2a_M}{\lambda_M} & -\frac{3\beta y_3}{\lambda_M} & -\frac{3\beta y_2}{\lambda_M} & -\frac{\gamma}{\lambda_M} \\ \frac{3\beta y_3}{\lambda_M} & -\frac{2a_M}{\lambda_M} & \frac{3\beta y_1}{\lambda_M} & \frac{\gamma}{\lambda_M} \\ \frac{12\beta y_1}{\lambda_M} & -\frac{12\beta y_2}{\lambda_M} & 0 & 0 \\ \frac{2\gamma}{\lambda_M} & -\frac{2\gamma}{\lambda_M} & 0 & 0 \end{pmatrix}. \quad (19)$$

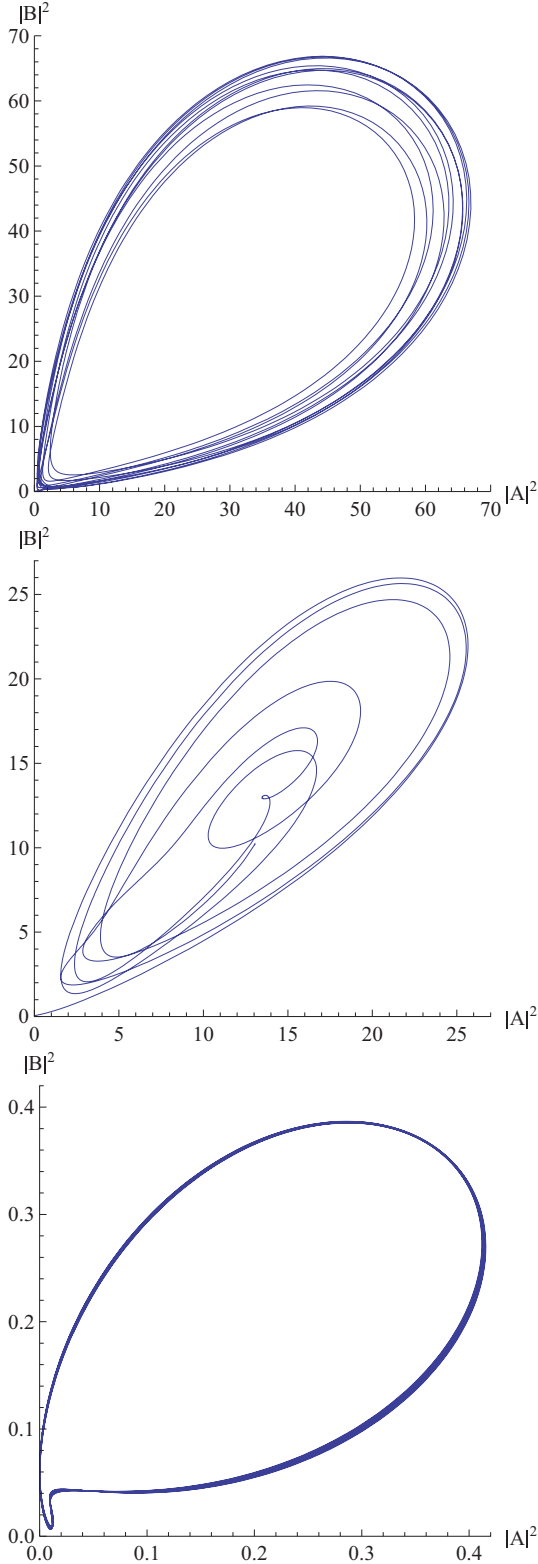

FIG. 6. (Color online) Numerical solutions $|A(x)|^2$ and $|B(x)|^2$ of (10) for $\varepsilon_0 = 0.1, a_M = -0.27, \lambda_M = -0.17,$ and $\beta = 0.002$. Top: $\gamma = 0.2$; middle: $\gamma = 0.27$; bottom: $\gamma = 0.4$.

One finds that the eigenvalues of the Jacobian matrix (19) at the equilibrium point (0,0,0,0) are $\lambda_{1,2} = 0$ and $\lambda_{3,4} = \pm \frac{2}{|\lambda_M|}\sqrt{a_M^2 - \gamma^2}$, corroborating that there is a bifurcation at $\gamma_c = |a_M|$ since the equilibrium point is a saddle when $0 < \gamma < |a_M|$, while it is a center when $\gamma > |a_M|$. The corresponding eigenvectors are $(\frac{\gamma}{2a_M}, \frac{\gamma}{2a_M}, 0, 1)^T$, $(0,0,1,0)^T$, and $(\frac{a_M \pm \sqrt{a_M^2 - \gamma^2}}{2\gamma}, \frac{\gamma}{2(a_M \pm \sqrt{a_M^2 - \gamma^2})}, 0, 1)^T$.

If we represent the homoclinic orbit as $A_h(x), B_h(x)$, then $u(x,t) \approx A_h(x)e^{it} + A_h^*(x)e^{-it}$, $v(x,t) \approx B_h(x)e^{it} + B_h^*(x)e^{-it}$ is an extended breather in the chain. Similar to the discrete system, the $O(\epsilon)$ correction produces a small pedestal and a second harmonic contribution. Clearly this pedestal will also go to zero in the wings if, as in [12], we insert purely lossy elements at both ends of the arrays.

We now turn our attention to the system at resonance (10). In this case one can see that for $\gamma = 0, |\epsilon_0| \ll 1$, the saddle point emanating from the origin $(A_E, B_E) \approx (\frac{i\epsilon_0}{a_M}, \frac{i\epsilon_0}{a_M})$ persists as the eigenvalues are $\lambda \approx \pm \frac{a_M}{\lambda_M}$. This is not surprising since we know saddle points are robust under perturbations. On the other hand, if $\gamma = O(1)$, the approximate critical point close to the origin $(A,B) \approx (0,0)$ is a center with eigenvalues $\lambda \approx \pm i\sqrt{\gamma^2 - a_M^2}$; thus a bifurcation must exist. For the more general case $\gamma, \epsilon_0 \neq 0$, rather than explicitly finding the bifurcation point, we show particular orbits at two different $\gamma$'s. The numerical solutions $|A(x)|$ and $|B(x)|$ are shown in Fig. 5, and the projection of the phase portrait in the plane $|A(x)|^2$ vs $|B(x)|^2$ is shown in Fig. 6. One can clearly observe the extended in $x$ quasiperiodic modes have a marked difference of reaching high maximum values for small $\gamma$ (top figures), while this is not the case for large $\gamma$ (bottom figures). Interestingly, the behavior at $\gamma = |a_M|$ shows a distinctively unique behavior.

As expected, at resonance the profile shown below represents to first order a spatially extended time-periodic nonlinear solution of the chain. This is so since forcing is extended throughout the chain. As stated above, the wings of this extended state can be made to go to zero if the elements of the chain at both ends are purely lossy.

Finally, the near-resonant case which needs to be analyzed as a full PDE problem with parametric forcing will not be considered here, but we expect to find regimes of quasiperiodic dynamics and chaotic regimes as well in similar dissipative forced systems.

## V. CONCLUSIONS

By developing a weakly nonlinear theory, restricting our analysis to the continuum model, we derived a dynamical system where, in the case of forcing away from resonance, we demonstrate the persistence for small $\mathcal{PT}$ parameter values $\gamma$ of a homoclinic orbit emerging from the conservative ($\gamma = 0$) case. This orbit represents a localized breather solution of the dimer chain. Our theory also proves that there is a critical value $\gamma_c$ above which the homoclinic orbit no longer exists, thus indicating a symmetry-breaking dynamics. We also show that at resonance a transition in the form of the stationary modes from large amplitude to small amplitude modes is also present.

Future work will consider studying the equivalent weakly nonlinear theory for discrete modes and a more detailed study of the time-dependent problem, including in particular the persistence of gap solitons for small $\gamma$ and small nonresonant, resonant, and near-resonant forcing.

**ACKNOWLEDGMENTS**

Work by A.B.A. was supported by the National Science Foundation, through the ECCS-1128593 IDR grant.